\documentclass[12pt]{iopart}
\usepackage{graphicx}
\usepackage{dcolumn}
\usepackage{bm}
\usepackage{epsf}

\begin{document}

\title{Multiphoton Bloch-Siegert shifts and level-splittings in
  spin-one systems}
\author{P L Hagelstein$^1$, I U Chaudhary$^2$}

\address{$^1$ Research Laboratory of Electronics, 
Massachusetts Institute of Technology, 
Cambridge, MA 02139,USA}
\ead{plh@mit.edu}

\address{$^2$ Research Laboratory of Electronics, 
Massachusetts Institute of Technology, 
Cambridge, MA 02139,USA}
\ead{irfanc@mit.edu}

\begin{abstract}
We consider a spin-boson model in which a spin 1 system is coupled to an 
oscillator.
A unitary transformation is applied which allows a separation of terms 
responsible for the Bloch-Siegert shift, and terms responsible for the level splittings at 
anticrossings associated with Bloch-Siegert resonances.
When the oscillator is highly excited, the system can maintain resonance for 
sequential multiphoton transitions.
At lower levels of excitation, resonance cannot be maintained because energy 
exchange with the oscillator changes the level shift.
An estimate for the critical excitation level of the oscillator is developed.
\end{abstract}

\pacs{32.80.Bx,32.60.+i,32.80.Rm,32.80.Wr}

\maketitle

\section{Introduction}
\label{sec:intro}

The interaction of a two-level system with oscillatory off-diagonal
 coupling leads to a shift in the transition energy, 
 known in the literature as the Bloch-Siegert shift
 \cite{BlochSiegert,Shirley}.   
When the shifted transition energy is resonant with an odd number of
oscillator quanta,  energy exchange between  
the two systems can occur.
This effect appears in energy level calculations as a level splitting at
the Bloch-Siegert resonances. 
Both the shift and splitting have been studied previously using models based on the
Rabi Hamiltonian (in which the oscillator 
is presumed classical)
\cite{BlochSiegert,Shirley,AhmadBullough,Fregenal,Forre,OstrovskyHorsdal}
and based on the spin-boson Hamiltonian (in which the oscillator is presumed 
quantum mechanical) \cite{Cohen,KlimovSainz}. 
It has been noted that the Bloch-Siegert shift arises from the
counter-rotating terms in the Hamiltonian, and  
that it disappears when the rotating-wave approximation is made
\cite{KlimovSainz,HattoriKobayashi} .

Our interest in this problem is motivated by the possibility of coherent energy 
exchange between quantum systems with highly mismatched characteristic energies.  
For example, we have been interested in the dynamics of energy exchange between a 
two-level system with a large transition energy $\Delta E$, and an oscillator with a 
small energy quantum $\hbar \omega_0$. 
The spin-boson Hamiltonian is one of the simplest models exhibiting such 
coherent energy exchange.
We recently considered the Bloch-Siegert shift \cite{HagelsteinChau1} and 
level splittings \cite{HagelsteinChau2} at the Bloch-Siegert resonances, the latter 
of which is a result of the coherent energy exchange between the two-level system and oscillator.

Most of the work cited above is focussed on the two-level problem, which is 
usually modeled as a spin 1/2 system interacting with a single harmonic oscillator. 
The generalization of the problem to systems involving higher spin leads to more 
complicated models which have not received comparable attention in the literature.
The Bloch-Siegert shift for the spin 1 case was studied previously by Hermann 
and Swain \cite{HermannSwain1,HermannSwain2}.  
Some progress has also been made in the case of the general spin problem
\cite{KlimovSainz}.

In this work we focus on the problem of a spin 1 system coupled to a simple 
harmonic oscillator.
Our earlier analysis in the spin 1/2 version of the problem made use of a 
unitary transform in which the shift and 
level splitting effects can be associated separately with different terms in
the rotated Hamiltonian. 
Such a separation serves as the basis for the development of analytic results 
for both shift and splitting which are useful over a wider range of coupling strength than available 
previously.
The unitary transform that we used for the spin 1/2 problem can be 
extended simply to higher spin models, allowing for a separation of shift and level splitting effects in 
more complicated problems.
We have decided to focus here on the spin 1 case since it provides a good 
example of this generalization.

Although the spin 1 system is a three-state system, we have found it useful to think about
it in terms of an underlying two-spin problem.  
For example, the Bloch-Siegert shift that we find below for the spin 1 can be understood simply
as arising from the individual shifts associated with two spin 1/2 systems.
Later on in this work, we find that the Bloch-Siegert resonance condition cannot be maintained at
modest $n$, which can be understood in terms of an initial resonant spin 1/2 transition that exchanges
energy with the oscillator, followed by a second spin 1/2 transition that is no longer resonant
since the oscillator is changed.

\section{Unitary equivalent Hamiltonian}
\label{sec:UnitaryEquivalent}

We focus on a spin 1 system coupled to an oscillator, using a spin-boson
Hamiltonian of the form

\begin{equation}
\hat{H} 
~=~
\Delta E {\hat{S}_z \over \hbar}
+
\hbar \omega_0 \hat{a}^\dagger \hat{a} 
+
U (\hat{a}^\dagger+\hat{a}) {2 \hat{S}_x \over \hbar}
\end{equation}
   
\noindent
We are interested in the regime where the photon excitation
$n_0$ is large, and where the transition energy is much greater than an
oscillator quantum ($\Delta E \gg \hbar \omega_0$).

As was the case in the spin 1/2 problem \cite{HagelsteinChau1,HagelsteinChau2}, 
it is useful to consider in the case of the spin 1 problem the unitary equivalent 
Hamiltonian

\begin{equation}
\hat{H}^\prime 
~=~ \hat{\mathcal{U}}^\dagger \hat{H} \hat{\mathcal{U}}
\end{equation}

\noindent
where 

\[
\hat{\mathcal{U}} 
~=~ 
\exp \left \lbrace - \frac{i}{2} 
          \arctan \left \lbrack
       {2 U (\hat{a} + \hat{a}^\dagger) \over \Delta E} 
                  \right \rbrack
       \frac{2 \hat{S}_y}{\hbar} \right \rbrace 
\]

\noindent 
The rotated Hamiltonian becomes

\begin{equation}
\hat{H}^\prime 
~=~
\hat{H}_0
~+~
\hat{V}
~+~
\hat{W}
\end{equation}

\noindent
where 

{\small

\begin{equation}
\hat{H}_0
~=~
\sqrt{\Delta E^2 + 4 U^2 (\hat{a}+\hat{a}^\dagger)^2 } {\hat{S}_z
  \over \hbar}
~+~
\hbar \omega_0 \hat{a}^\dagger \hat{a} 
\end{equation}

\begin{samepage}

\[
\hat{V} = 
{i \hbar \omega_0 \over 2}
\left \lbrace
\left [
{ \displaystyle{  U  \over \Delta E}  \over 1+  \left [ \displaystyle{ 2U 
(\hat{a}+\hat{a}^\dagger) \over \Delta E} \right ]^2  }
\right ]
(\hat{a}-\hat{a}^\dagger)
\right .
\]
\begin{equation}
\ \ \ \ \ \ \ \ \ \ \ \ \ \ \ \ \ \ \ \ \ \ \ \ \ \ \ \ \ \ \ \ \ \ \ \
\left .
~+~
(\hat{a}-\hat{a}^\dagger)
\left [
{ \displaystyle{  U  \over \Delta E}  \over 1+   \left [ \displaystyle{ 2 U 
(\hat{a}+\hat{a}^\dagger) \over \Delta E} \right ]^2  }
\right ]
\right \rbrace
\frac{2 \hat{S}_y}{\hbar} 
\end{equation}

\end{samepage}

\begin{equation}
\hat{W} = \hbar \omega_0 
\left \lbrace
{ \displaystyle{  U  \over \Delta E}  \over 1+   \left [ \displaystyle{
      2 U (\hat{a}+\hat{a}^\dagger) \over \Delta E} \right ]^2  } 
 \frac{2 \hat{S}_y}{\hbar} \right \rbrace^2
\end{equation}

}
\noindent 
Within the parameter space of interest to us, the $\hat{W}$ term is 
small; consequently, we neglect it in what follows.

\section{Approximate energy eigenvalues}
\label{sec:EstimateEnergy}

In the large $n$ limit, the oscillator has a strong impact on the spin system, 
but the spin system impacts the oscillator only weakly.
As a result, the energy levels are found to be reasonably well approximated away 
from the level anticrossings by

\begin{equation}
\label{eq:approximation}
E_{n,M}(g) 
~=~ 
\Delta E(g)M
+
\hbar \omega_0 n 
\end{equation}

\noindent
where $\Delta E(g)$ is the dressed two-level system energy. 
This is very much like the behaviour found for the spin 1/2 version of the 
problem discussed previously \cite{HagelsteinChau1}.
It is useful to adopt the same definition for the dimensionless coupling 
constant $g$ for the spin 1 version of the problem

\[
g  ~=~  \frac{U \sqrt{n_0}}{\Delta E} 
\]

\noindent
that we used before for the spin 1/2 case.

The dressed two-level system energy $\Delta E(g)$ is also the same, as would be 
expected since it comes about from the same basic interaction. 
We can think of the spin 1 system as being made up of two spin 1/2 systems 
coupled identically to the same oscillator. 
Since the Bloch-Siegert shift is the result of the interaction between a single 
two-level system and oscillator, 
we should expect that it will be nearly the same per two-level system if there 
are more than one.

We can make use of the rotation to see this.
Consider eigenfunctions of the unperturbed $\hat{H}_0$ part of the rotated 
Hamiltonian

\begin{equation}
E_{n,M} \psi_{n,M}
~=~
\hat{H}_0 \psi_{n,M} 
\end{equation}

\noindent
We can separate variables (as done in the spin 1/2 case) to write

\[
\psi_{n,M} ~=~ |S,M \rangle u_{n,M}(y)
\]

\noindent 
where $u_{n,M}(y)$ satisfies 

{\small

\begin{equation}
\left ( E_{n,M} + {\hbar \omega_0 \over 2} \right ) u_{n,M}
~=~ 
{\hbar \omega_0 \over 2} \left [-{d^2 \over dy^2} + y^2 \right ]u_{n,M}
+
M \sqrt{\Delta E^2 + 8 U^2 y^2 } u_{n,M}
\end{equation}

}

\noindent
In the large $n$ limit, this is just a modified simple harmonic oscillator with 
a small $M$-dependent perturbation.  
By adopting a simple harmonic oscillator wavefunction as a trial solution in a 
variational computation, 
we obtain approximate energy levels of the form of Equation (\ref{eq:approximation}) 
where

\begin{equation}
\label{eq:DEApprox}
\Delta E(g)  
~=~  
\left \langle n \left |
\sqrt{ 1 + {4 U^2  (\hat{a} + \hat{a}^\dagger)^2 \over \Delta E^2 }}
\right | n  \right \rangle   
\end{equation}

\noindent
This dressed transition energy, identical to what we found in the spin 1/2
case \cite{HagelsteinChau1}, is in good agreement with the numerical
results away from the level anticrossings.

\section{Multiphoton resonances}
\label{sec:MultiphotonResonances}

As mentioned above, our interest in this model is driven in part by the 
possibility of coherent energy exchange
between systems with strongly mismatched characteristic energies.
In the spin 1 problem, resonances occur associated with anticrossings when the 
dressed two-level system energy $\Delta E(g)$ matches an odd number of oscillator quanta.
In the large $n$ limit, the oscillator is impacted only weakly through the 
exchange of a modest
number of oscillator quanta; in which case the dressed energy $\Delta E(g)$ is 
approximately invariant, and
we are able to develop a resonance condition applicable to both transitions. 
When $n$ is smaller, the change in the number of oscillator quanta may be a 
significant fraction
of the total number of oscillator quanta; in which case the dressed two-level 
transition energy
may be on resonance for one transition and off resonance for the other.
We begin our discussion here focusing on the first case, since it is simpler.

\subsection{Large $n$ resonance condition}

We are interested then in the resonance condition at which the dressed
transition energy is matched to an odd number of oscillator quanta.  
To proceed, we consider three basis states which are eigenstates of the rotated 
Hamiltonian $\hat{H}_0$:

{\small

\begin{equation}
\label{eq:BasisStates}
\phi_{-1} ~=~ u_{n+\Delta n,-1}(y) |S,-1 \rangle 
\ \ \ \ \
\phi_{ 0} ~=~ u_{n,0}(y) |S, 0 \rangle 
\ \ \ \ \
\phi_{ 1} ~=~ u_{n-\Delta n,1}(y) |S, 1 \rangle 
\end{equation}

}

\noindent
where $S=1$.  
These states have energies $\epsilon_{-1}$, $\epsilon_0$ and $\epsilon_1$
respectively, which in the large $n$ limit are given approximately by

\begin{equation}
\epsilon_{-1} 
~=~ 
\hbar \omega_0 \left ( n + \Delta n  \right ) - \Delta E(g) 
\end{equation}
\begin{equation}
\epsilon_{0} 
~=~ 
\hbar \omega_0  n 
\end{equation}
\begin{equation}
\epsilon_{ 1} 
~=~ 
\hbar \omega_0 \left ( n - \Delta n  \right ) + \Delta E(g) 
\end{equation}

\noindent
with $\Delta E(g)$ taken to be a constant for both transitions.
The resonance condition in this case is

\begin{equation}
\Delta E(g) ~=~ \Delta n \hbar \omega_0
\end{equation}

\noindent
where, as in the two-level case,  the number of oscillator quanta exchanged 
$\Delta n$ must be odd for 
level splitting to occur.

\subsection{Comparison with previous work}

As mentioned above, Hermann and Swain \cite{HermannSwain2} have
calculated the Bloch-Siegert shift for the spin 1 case.   
In their notation, the resonance condition to fourth order is

\begin{equation}
\omega^{min} 
~=~ 
\frac{\omega_0}{p} + \frac{2 V^2 p}{(p^2 - 1) \omega_0} 
-
\frac{p^3(3p^2 - 7)V^4}{(p^2 - 1)^3 \omega_0^3} 
\end{equation}

\noindent 
In order to write this resonance condition in our notation, we make the 
following replacements 

\[
V \rightarrow U \sqrt{2n}, \; \; \; \;
\omega^{min} \rightarrow \omega_0, \; \; \; \;
\omega_0 \rightarrow \Delta E, \; \; \; \; 
p \rightarrow (2k + 1)
\] 

\noindent 
to obtain 

{\small

\begin{equation}
\label{eq:ResonanceSwain}
\Delta E \left[ 1 + \frac{4 g^2 (2k+1)^2}{4k(k+1)} - \frac{4 (2k+1)^4
  \left[3 (2k+1)^2  - 7 \right]g^4}{ \left[4k(k+1)\right]^3 } + \cdots
  \right ] 
~=~  (2k + 1) \hbar \omega_0
\end{equation}
}

%
%
%
%
\noindent Note that
Equation~(\ref{eq:ResonanceSwain}) is exactly the same condition 
as obtained by Ahmad and Bullough \cite{AhmadBullough, HagelsteinChau1}
for the spin 1/2 problem.  

To compare our calculation to these perturbative results, we can expand
$\Delta  
E(g)$ from Equation (\ref{eq:DEApprox}) in powers of $g$
to obtain the condition for the $(2k + 1)$th resonance as 

\begin{equation}
\label{eq:ResonanceCondition}
\Delta E \left(1 + 4 g^2 - 12 g^4 \cdots \right) 
~=~ 
(2k + 1) \hbar \omega_0
\end{equation}

\noindent
where $\Delta n = 2k+1$.   
This condition is the same as was found in the spin 1/2 problem
\cite{HagelsteinChau1}. 
In the limit that $k \gg 1$, we see that
Equation~(\ref{eq:ResonanceSwain}) reduces to our result   
[Equation~(\ref{eq:ResonanceCondition})].

\section{Dynamics: resonant case}
\label{sec:Dynamics}

The basis states that we chose [Equation (\ref{eq:BasisStates})] are 
matched total energy states; 
the spin 1 system pieces are combined with modified oscillator states 
in which an odd number of oscillator quanta are matched to the dressed two-level transition energy.
When the coupled system makes transitions between these different basis states, 
the resulting
dynamics describe coherent energy exchange between component quantum systems with very 
different characteristic energies.
The resonant case in the high $n$ limit is most interesting in this regard,
since both two-level systems make transitions, and a full $2 \Delta n$ oscillator
quanta are exchanged.

\subsection{Three-state model}

To proceed, we consider a dynamical state $\psi(t)$ constructed from the three 
basis states 
listed in Equation (\ref{eq:BasisStates})

\begin{equation}
\psi(t) ~=~ c_{-1}(t) \phi_{-1} + c_0(t) \phi_0 + c_1(t) \phi_1
\end{equation}

\noindent
In the large $n$ limit, these three basis states become degenerate. 
In addition, the different matrix elements become roughly equal

\begin{equation}
\langle \phi_{-1}|\hat{V}|\phi_{ 0} \rangle 
~\approx~ 
\langle \phi_{ 0}|\hat{V}|\phi_{ 1} \rangle 
~\to~
 v
\end{equation}
\noindent 
where 

\begin{equation}
v 
~=~   
\left [
{ 2 \hbar \omega_0 U  \over \Delta E }
\right ]
I
\end{equation}
\noindent and 

\begin{equation}
\int_{-\infty}^\infty
u_{n+\Delta n,M}(y) 
{ 1  
\over 1+ 8 \left ( \displaystyle{ U y \over \Delta E} \right )^2  
}
\left (
{d \over dy}
u_{n,M}(y)
\right )
dy
~\to~ I
\label{Ieqn}
\end{equation}

\begin{equation}
-\int_{-\infty}^\infty
\left (
{d \over dy}u_{n+\Delta n,M}(y)
\right )
{ 1  
\over 1+ 8 \left ( \displaystyle{ U y \over \Delta E} \right )^2  
}
u_{n,M}(y)
dy
~\to~ I
\label{Jeqn}
\end{equation}

\noindent 
The two integrals appearing here have been found to approach a common limit 
(to within the sign) based on calculations using numerical solutions for the
rotated frame $\hat{H}_0$ problem, and based on calculations using the WKB approximation.
This behaviour can be understood simply by noting that the the derivative
can be expressed in terms of raising and lowering operators and
in the large $n$ limit the $u_{n}(y)$ functions 
are very nearly pure harmonic oscillator states. 
%

The  inclusion of the perturbation $\hat{V}$ leads to the restricted
Schr\"odinger equation 

\begin{equation}
\label{eq:Dynamics}
i \hbar {d \over dt}
\left (
\begin{array} {c}
c_{-1} \cr
c_{0}  \cr
c_{1}  \cr
\end{array}
\right )
~=~
\left (
\begin{array} {ccc}
\epsilon_0 & v   & 0    \cr
v & \epsilon_0 & v    \cr
0   & v & \epsilon_0  \cr
\end{array}
\right )
\left (
\begin{array} {c}
c_{-1} \cr
c_{0}  \cr
c_{1}  \cr
\end{array}
\right )
\end{equation}


\subsection{Dynamical solution}

\epsfxsize = 3.80in
\epsfysize = 2.50in
\begin{figure} [t]
\begin{center}
\mbox{\epsfbox{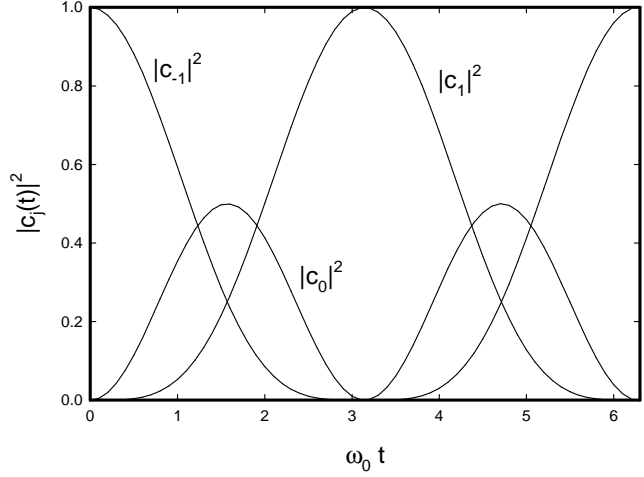}}
\caption{Occupation probabilities as a function of time for the
  degenerate three state model. 
The time axis is in units of $\omega_0 t$ where $\omega_0 = \sqrt{2} v 
/\hbar$.}  
\label{Fig34x1}
\end{center}
\end{figure}

Solutions for Equation~(\ref{eq:Dynamics}) can be readily constructed using an 
eigenfunction expansion.  
One interesting solution is

\begin{equation}
\label{eq:ThreeStateSoln}
\left (
\begin{array} {c}
c_{-1}(t) \cr
 \cr
c_{0}(t)  \cr
 \cr
c_{1}(t)  \cr
\end{array}
\right )
~=~
e^{-i \epsilon_0 t / \hbar}
\left (
\begin{array} {c}
{1 \over 2} \left [ \cos {\sqrt{2} v t \over \hbar} + 1 \right ]       \cr
  \cr
{i \over \sqrt{2}} \sin {\sqrt{2} v t \over \hbar}  \cr
  \cr
{1 \over 2} \left [  \cos {\sqrt{2} v t \over \hbar} - 1 \right ]       \cr
\end{array}
\right )
\end{equation}

\noindent
The associated probabilities are illustrated in Figure~\ref{Fig34x1}.
One sees that the system starts in state $\phi_{-1}$ with unity probability, 
moves through
state $\phi_0$, and then reaches $\phi_{1}$ with probability unity.  
The solutions are periodic, so that the system cycles back and forth between the 
different states.

\subsection{Evolution of expectation values}

\begin{figure} [t]
\begin{center}
\mbox{\epsfbox{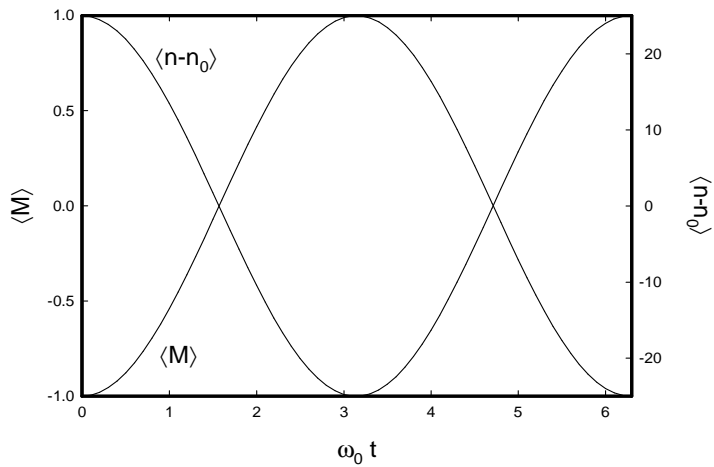}}
\caption{Expectation value $\langle M \rangle$ and $\langle n - n_0
  \rangle$ as a function of time. 
The time axis is in units of $\omega_0 t$ where $\omega_0 = \sqrt{2} v
/\hbar$.  This computation is for a resonant energy exchange process where 25
oscillator quanta are exchanged for one dressed two-level system quantum.
} 
\label{Fig34x2}
\end{center}
\end{figure}

In Figure~\ref{Fig34x2} we show results for the expectation values $\langle M 
\rangle$ and 
$\langle n - n_0 \rangle$ in a resonant energy exchange process where 25
oscillator quanta are exchanged for a single dressed two-level system quantum.  
The expectation values in this case are computed according to

\begin{equation}
\langle M \rangle = \sum_{M^\prime} M^\prime |c_{M^\prime}(t)|^2
\ \ \ \ \ \ \ \ \ \
\langle n-n_0 \rangle = - \sum_{M^\prime} M^\prime \Delta n 
|c_{M^\prime}(t)|^2 
\end{equation}

\noindent
One sees in this case a complete energy exchange between the two systems, where 
the excitation 
energy of two dressed two-level systems [$2 \Delta E(g)]$ is exchanged for an 
equal amount 
of oscillator energy ($2 \Delta n \hbar \omega_0$).

\epsfxsize = 3.80in
\epsfysize = 2.50in

\subsection{Oscillations of $\langle M \rangle$ are sinusoidal}

We can look at the dynamics of energy exchange in the large $n$ resonance case
in another way.
Within the set of basis states that we have selected, the basis states are 
degenerate
on resonance, and the coupling is proportional to matrix elements of 
$\hat{S}_x$ [see Equation~(\ref{eq:Dynamics})].
In this case, we can determine the evolution of $\langle \hat{S}_z \rangle$ for
the spin 1 case (or in other cases as well) through commutation with the 
interaction term of the associated restricted Hamiltonian.
The result is

\begin{equation}
{d^2 \over dt^2} 
\langle \hat{S}_z \rangle
~=~
- \Omega^2 \langle \hat{S}_z \rangle 
\end{equation}

\noindent
where the frequency $\Omega$ is

\begin{equation}
\Omega ~=~ {2\sqrt{2}  \omega_0 U I \over \Delta E}
\end{equation}

\noindent
Hence if all the two-level systems are initially in the ground state (in the 
rotated frame), 
in general the average $\langle \hat{S}_z \rangle$ will exhibit sinusoidal dynamics 
according to

\begin{equation}
\langle \hat{S}_z \rangle 
~=~
\langle M \rangle 
~=~
- S \cos (\Omega t)
\end{equation}

\noindent
This is consistent with the results shown in Figure 2. Note
that these dynamics are in the rotated frame, and could be obtained if
the interaction $U$ were turned on adiabatically.

\section{Finite $n$ effects}
\label{sec:FiniteN}

Up to this point we have assumed that $n$ is ``large enough'' so that the basis 
states are degenerate.  
However,  when $n$ is not overly large, the levels are not degenerate and hence 
it is not possible to arrange for a clean energy exchange as described above.  
This motivates an interest in understanding how large $n$ must be so that the 
system acts as if it is degenerate.

\subsection{Parameterization of basis state energies}

For finite $n$, we have found (based on numerical calculations) that the energy 
levels of the original Hamiltonian $\hat{H}$ 
can be accurately fit away from resonance using the expansion of the form

\begin{equation}
E_{n,M}
~=~
A + B(n-n_0) + C M  +
D M (n-n_0) +
F M^2 + \cdots
\end{equation}

\noindent
where $A$, $B$, $C$, $D$, and $F$ are fitting coefficients.  The terms
quadratic in $\Delta n$ are not that significant because the oscillator
is highly excited. 
We can relate the basis state energies $\epsilon_{-1}$ and $\epsilon_{1}$ to $\epsilon_0$ using this parameterization
to obtain

\begin{equation}
\epsilon_{-1} ~=~ \epsilon_0 + B \Delta n - C - D \Delta n + F + \cdots
\end{equation}

\begin{equation}
\epsilon_{ 1} ~=~ \epsilon_0 - B \Delta n + C - D \Delta n + F + \cdots
\end{equation}

\noindent
We take the approximate resonance condition to be

\begin{equation}
B \Delta n ~=~ C
\end{equation}

\noindent
which is nearly equivalent to our resonance condition from above

\begin{equation}
\Delta n \hbar \omega_0 ~=~ \Delta E(g)
\end{equation}

\subsection{Mismatch in basis state energies at resonance}

When this resonance condition is satisfied, the energy levels (in the absence of 
coupling terms) 
are still not matched.  Instead, we find that $\epsilon_0$ lies above $\epsilon_1$ and 
$\epsilon_{-1}$ due
to the presence of higher-order terms in the fitting expansion

\begin{equation}
\epsilon_{-1} ~=~ \epsilon_0  - D \Delta n + F + \cdots
\end{equation}

\begin{equation}
\epsilon_{ 1} ~=~ \epsilon_0  - D \Delta n + F + \cdots
\end{equation}

\noindent
Both analytic and numeric computations lead to the conclusion that

\begin{equation}
F - D \Delta n  ~\to~ { {\rm constant} \over n_0}
\end{equation}

\noindent
for large $n_0$.  
The numerator on the RHS is a constant that can be determined from 
parameterizing the levels 
in a direct numerical calculation, or can be estimated from perturbation theory 
in the rotated frame.
For the purposes of discussion, we may write this as

\begin{equation}
F - D \Delta n 
~\to~ 
{ [n_0(F - D \Delta n)]_{n_0=\infty}  
 \over n_0}
\end{equation}

\subsection{Determination of critical $n$ where splitting matches coupling}

The matrix element that produces transitions in the rotated frame can be
approximated by

\begin{equation}
\langle \phi_0 | \hat{V} | \phi_1 \rangle
~=~
\frac{2 \hbar \omega_0 U}{\Delta E} I
~\to~
 2 \hbar \omega_0
g 
\left [ {I \over \sqrt{n_0}} \right ]_{n_0=\infty} 
\end{equation}

\noindent
We can now estimate the number of oscillator quanta $n$ required to make the
coupling matrix element equal in magnitude to the basis state splitting by 
requiring

\begin{equation}
|E_1-E_0| 
~=~
{ [n_0(F - D \Delta n)]_{n_0=\infty}  
 \over n_0}
~=~ 
\sqrt{2} |\langle \phi_0 | \hat{V} | \phi_1 \rangle|
\end{equation}

\noindent
The $\sqrt{2}$ that appears here reflects the extra factor in the 
level splittings obtained 
from a diagonalization of the three-state model on resonance.  
This is satisfied when $n$ is equal to a critical value

\begin{equation}
\label{eq:Ncrit}
n_{crit} 
~=~ 
\left |
{
[n_0(F - D \Delta n)]_{n_0=\infty}  
\over
2 \sqrt{2} \hbar \omega_0
g 
\left [ {I / \sqrt{n_0}} \right ]_{n_0=\infty} 
}
\right |
\end{equation}

\epsfxsize = 3.80in
\epsfysize = 2.50in
\begin{figure} [t]
\begin{center}
\mbox{\epsfbox{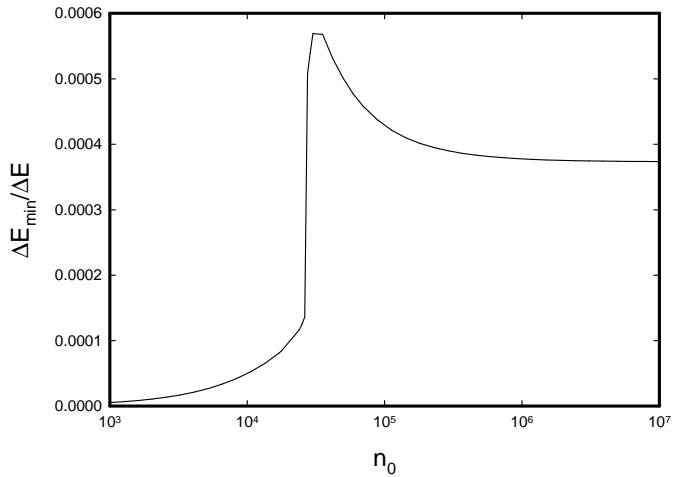}}
\caption{Level splitting from direct solution of original Hamiltonian $\hat{H}$ 
problem
as a function of $n_0$.  The level splitting is between states that in the 
rotated 
frame are mostly composed of $M=-1$, $\Delta n = 15$, and $M=1$, $\Delta n = 
-15$, for
a model in which $\Delta E=11 \hbar \omega_0$.
} 
\end{center}
\label{Fig34x3}
\end{figure}

\subsection{Numerical example}

In Figure 3 is shown the level splittings for state computed from the original
Hamiltonian $\hat{H}$ without rotation in an example that illustrates this 
effect.  
In this calculation, we have selected a model with $\Delta E = 11 ~\hbar 
\omega_0$, and we
focus on the resonance at $\Delta E(g) = 15~ \hbar \omega_0$, where $g$ is 
determined from numerical optimization to minimize the level splitting at each $n_0$.  
The numerical data shows a resonance behaviour similar to what we would expect 
from the three-state model above.  
At large $n_0$, the basis states are nearly degenerate relative to the
coupling matrix element, and we find maximum level splitting.  
At small $n_0$, the basis states are separated by more than the coupling matrix 
element, so that the second-order
coupling between $\phi_{-1}$ and $\phi_1$ is small, and hence the level 
splitting is also small.  
The critical number of oscillator quanta in this case is about $2.7 \times 
10^4$.

\subsection{Estimation of $n_{crit}$ from analytic estimate}

For this problem, the quantity $I/\sqrt{n}$ is found from a WKB calculation to 
be

\begin{equation}
\left [ {I \over \sqrt{n_0}} \right ]_{n_0=\infty} 
~=~
1.77 \times 10^{-3}
\end{equation}

\noindent
The $D$ term estimated from

\begin{equation}
D ~=~ \left ( {\partial^2 \over \partial n \partial M} E_{n,M} \right 
)_{n=n_0,M=0}
\end{equation}

\noindent
is used to obtain

\begin{equation}
n_0D ~\to~ 6.60 ~\left ( {1 \over 2} \hbar \omega_0 \right )
\end{equation}

\noindent
as $n_0$ goes to $\infty$.
From a direct parameterization of the energy levels we obtain

\begin{equation}
n_0F ~\to~ -10.4 ~ \left ( {1 \over 2} \hbar \omega_0 \right )
\end{equation}

\noindent
We can compare this result from the value obtained using the WKB approximation 

\begin{equation}
n_0 F 
~=~ 
{n_0 \over 2} \left ( {\partial^2 \over \partial M^2} E_{n,M} \right )
~=~
-11.0~ \left ( {1 \over 2} \hbar \omega_0 \right )
\end{equation}

\noindent
These parameters are combined to produce the estimate based on Equation~(\ref{eq:Ncrit}) 

\begin{equation}
n_{crit} ~=~ 3.26 \times 10^4
\end{equation}

\noindent
This estimate is in reasonable agreement with the calculation of Figure 3. 
We conclude that a simple three-state model provides a good foundation for 
estimating 
how large $n$ must be in order for the basis states to be degenerate relative to 
the coupling.

\section{Discussion}

We have previously applied a rotation in the case of the spin 1/2 problem in 
order to produce
a dressed system in which the unperturbed Hamiltonian $\hat{H}_0$ provides a 
good approximation
away from the level anticrossings at resonance \cite{HagelsteinChau1}, and where 
the perturbation $\hat{V}$ gives most
of the coupling responsible for the level splittings at the anticrossings 
\cite{HagelsteinChau2}.
Here we have applied a similar rotation to the spin 1 problem, in which case a 
similar separation
of the rotated Hamiltonian occurs.
Although our discussion here is focused on the spin 1 problem, we have found 
that similar good
results are also obtained in the case of higher spin as well.
We find that the Bloch-Siegert shift in the spin 1 case is close to that of the 
spin 1/2 problem,
in agreement with the perturbative result of Hermann and Swain 
\cite{HermannSwain2}.

We have considered level splitting at the anticrossing in association with the 
system dynamics
at resonance.
In the large $n$ limit, energy exchange between the spin 1 system and oscillator 
constitutes only a 
small perturbation to the oscillator, so that the dressed transition energy is 
not changed, and
that the same resonance condition applies to both transitions.
In this case, it is possible for complete energy exchange to occur between the 
spin 1 system and the oscillator.
We have given an analytic solution for the dynamics in the case of resonance.

At more modest $n$, the oscillator is modified sufficiently by a change of 
$\Delta n$ oscillator
quanta so that the resonance condition no longer holds for a second transition.
In this case, we might think of the coupled system as being made up of two spin 
1/2 systems with 
different dressed transition energies weakly coupled to the oscillator.
Accordingly, at larger $n$ we might think of the coupled system as a dressed 
spin 1 system
weakly coupled to the oscillator.
The transition between these two kinds of behaviour is determined by a critical 
excitation 
of the oscillator.
We have developed an analytic estimate for this critical number of oscillator 
quanta $n_{crit}$ [see Equation~(\ref{eq:Ncrit})]
in terms of fitting parameters which we can derive directly from solutions of 
the unrotated
Hamiltonian $\hat{H}$, from numerical solutions of the unperturbed part of the 
rotated Hamiltonian $\hat{H}_0$,
or from the WKB approximation as applied to the rotated problem.

\end{document}